\documentclass[aps,10pt,prl,floatfix,twocolumn,nofootinbib,superscriptaddress,groupedaddress,showkeys]{revtex4-2}

\usepackage{amsmath,amssymb,bm,color,float,graphicx,mathrsfs}
\usepackage{hyperref}
\usepackage{Macro}
\hypersetup{colorlinks=true,linkcolor=blue,citecolor=blue,urlcolor=blue}

\newcommand{\compile}{}
\newcommand{\submitletter}{}

\begin{document}

\ifdefined\compile

\title{Resolving the negative effective neutrino mass parameter with cosmic birefringence}

\author{Toshiya Namikawa}
\affiliation{Department of Applied Mathematics and Theoretical Physics, University of Cambridge, Wilberforce Road, Cambridge CB3 0WA, United Kingdom}
\affiliation{Center for Data-Driven Discovery, Kavli IPMU (WPI), UTIAS, The University of Tokyo, Kashiwa, 277-8583, Japan}
\affiliation{Kavli Institute for Cosmology, University of Cambridge, Madingley Road, Cambridge CB3 OHA, United Kingdom}

\date{\today}

\begin{abstract}
The recent measurement of baryonic acoustic oscillations by the Dark Energy Spectroscopic Instrument reveals a mild tension with observations of the cosmic microwave background (CMB) within the standard $\Lambda$CDM cosmological model. This discrepancy leads to a preference for a total neutrino mass that is lower than the minimum value inferred from neutrino oscillation experiments. Alternatively, this tension can be eased within $\Lambda$CDM by assuming a higher optical depth ($\tau \simeq 0.09$), but such a value conflicts with large-scale CMB polarization data. We point out that cosmic birefringence, as suggested by recent Planck reanalyses, resolves this discrepancy if the birefringence angle varies significantly during reionization. Specifically, we consider the fact that the measured cosmic birefringence angle $\beta_0=0.34\pm0.09\,(1\,\sigma)\,$deg has the phase ambiguity, i.e., the measured rotation angle is described by $\beta=\beta_0+180n\,$deg ($n\in \mathbb{Z}$). We show that cosmic birefringence induced by axion-like particles with nonzero $n$ suppresses the reionization bump, allowing a higher $\tau$ consistent with data. We provide a viable parameter space where the birefringence effect simultaneously accounts for the low-$\ell$ polarization spectra, the Planck $EB$ correlations, and the elevated value of $\tau$, suggesting a key role for cosmic birefringence in current cosmological tensions.
\end{abstract} 



\maketitle

\fi

\ifdefined \submitletter
  {\it Introduction---}
\else
  \section{Introduction} \label{sec:intro}
\fi
Recent measurements of the cosmic microwave background (CMB) and baryon acoustic oscillations (BAO) reveal a mild tension within the standard spatially-flat $\Lambda$ cold dark matter ($\Lambda$CDM) framework~\cite{DESI:2025:DR2,SPT-3G:2025}. This discrepancy leads to a preferred value for the sum of neutrino masses that is lower than the minimal mass required by neutrino oscillation experiments~\cite{DESI:2025:neutrino}. When an effective neutrino mass parameter that allows for negative values is introduced, the minimal neutrino mass is excluded at a significance level of approximately $3\,\sigma$~\cite{Craig:2024:neutrino,Green:2024:neutrino,Elbers:2024:neutrino,Lynch:2025:neutrino-tension,DESI:2025:neutrino}, although the exact statistical significance depends on the Planck likelihood~\cite{Herold:2024:mnu,Naredo-Tuero:2024:mnu}.

Within the standard cosmological framework, a proposed resolution of this tension involves reinterpreting the discrepancy as a shift in the CMB optical depth, $\tau$, as constrained by Planck~\cite{Giare:2023:tau,Loverde:2024:mnu,Sailer:2025:tau,Jhaveri:2025:tau}. By disregarding the low-$\l$ polarization data, the discrepancy can be reconciled by assuming a higher value of $\tau$ than that obtained from the full Planck analysis. For example, a $\Lambda$CDM model with $\tau=0.09$ is consistent with the minimum mass of the normal hierarchy model at $0.9\,\sigma$ \cite{Sailer:2025:tau}. A higher value of $\tau = 0.09$ also reduces the preference for the $w_0w_a$ dynamical dark energy model over the $\Lambda$CDM model to $1.4\,\sigma$ \cite{Sailer:2025:tau}, and also diminishes the significance of both the lensing and curvature anomalies \cite{Giare:2023:tau}. However, such a higher $\tau$ is inconsistent with the Planck low-$\l$ measurement at $3.8\,\sigma$ significance \cite{Sailer:2025:tau}. Although the constraint on $\tau$ from the low-$\l$ $E$-mode power spectrum, in principle, depends on the assumed reionization history~\cite{Mortonson:2007:reion,Jiang:2024}, the constraint on $\tau$ from the current Planck data is robust against the reionization history~\cite{Ilic:2025:tau}, and uncertainties in the reionization model do not significantly alleviate the tension~\cite{Jhaveri:2025:tau}. Thus, increasing $\tau$ within the standard assumptions has not succeeded in resolving the discrepancy. Setting aside the possibility that residual systematics remain in the low-$\l$ $E$-mode polarization, this may point toward new physics beyond the $\Lambda$CDM model.

Another intriguing possibility for the beyond-standard cosmology arises from recent reanalyses of polarization data from the Planck and Wilkinson Microwave Anisotropy Probe (WMAP), which suggest the presence of cosmic birefringence---a phenomenon in which the polarization plane of light rotates as it propagates through space~\cite{Minami:2020:biref,Diego-Palazuelos:2022,Eskilt:2022:biref-freq,Eskilt:2022:biref-const,Eskilt:2023:EDE} (see Refs.~\cite{Alighieri:2015:review,Komatsu:2022:review} for review). The latest measurement by Ref.~\cite{Eskilt:2022:biref-const} shows a preference for isotropic cosmic birefringence at $3.6\,\sigma$ significance. 
As a parity-violating phenomenon, cosmic birefringence presents a promising signal of new physics beyond both the $\Lambda$CDM model and the Standard Model of particle physics~\cite{Nakai:2023}.

Cosmic birefringence can arise from a pseudoscalar field $\phi$, such as axion-like particles (ALPs), coupled to the electromagnetic field via a Chern–Simons interaction:
\al{
    \mathcal{L}\supset -\frac{1}{4}g\phi F_{\mu\nu}\tilde{F}^{\mu\nu}
    \,,
} 
where $g$ is the coupling constant, $F_{\mu\nu}$ is the electromagnetic field tensor, and $\tilde{F}^{\mu\nu}$ is its dual. This mechanism has been widely explored in various cosmological contexts, including ALPs as dark energy candidates~\cite{Carroll:1998:DE,Liu:2006:biref-time-evolve,Panda:2010,Fujita:2020:forecast,Fujita:2020:biref,Choi:2021aze,Obata:2021,Gasparotto:2022uqo,Galaverni:2023}, early dark energy scenarios~\cite{Fujita:2020:biref,Murai:2022:EDE,Eskilt:2023:EDE,Kochappan:2024:biref}, and axion dark matter models~\cite{Finelli:2009,Sigl:2018:biref-sup,Liu:2016:AxionDM,Fedderke:2019:biref,Zhang:2024dmi,Namikawa:2025:Planck}. Additional sources proposed include topological defects~\cite{Takahashi:2020tqv,Kitajima:2022jzz,Jain:2022jrp,Gonzalez:2022mcx,Lee:2025:biref-DW} and possible quantum gravity effects~\cite{Myers:2003fd,Balaji:2003sw,Arvanitaki:2009fg}.

In this Letter, we point out that cosmic birefringence resolves the ``$\tau$ tension'' if the birefringence angle varies significantly during the epoch of reionization. A large variation of the rotation angle causes the polarization signals generated during reionization to acquire different phases along the line of sight. When these rotated signals are integrated, the reionization signals are partially canceled due to different phases, leading to a suppression of the reionization bump in the $E$-mode power spectrum. Consequently, a larger value of $\tau$ than that obtained in the standard $\Lambda$CDM model can be accommodated, thereby explaining the low-$\l$ $E$-mode power spectrum. To illustrate this effect, we focus on an ``$n\pi$-phase ambiguity'' in the recent birefringence measurement of $\beta_0=0.34\pm0.09\, (1\,\sigma)\,$deg---i.e., the rotation angle that fits the data is degenerate with $\beta_0+180n\,$deg for $n\in\mathbb{Z}$~\cite{Naokawa:2024xhn}. We demonstrate that if cosmic birefringence is induced by ALPs with $n\not=0$, it can accommodate a larger value of $\tau$, thereby alleviating the tension between the CMB and Dark Energy Spectroscopic Instrument (DESI) results, while simultaneously being consistent with the birefringence angle constrained by the $EB$ power spectrum.

\ifdefined \submitletter
  {\it CMB angular power spectra with cosmic birefringence---}
\else
  \section{CMB angular power spectra with cosmic birefringence} \label{sec:theory}
\fi
Let us introduce the CMB angular power spectra induced by cosmic birefringence from ALPs~\cite{Liu:2006:biref-time-evolve,Finelli:2009,Gubitosi:2014:biref-time,Lee:2013:biref,Sherwin:2021:biref,Nakatsuka:2022,Murai:2022:EDE,Naokawa:2023,Murai:2024yul}. 
When cosmic birefringence is sourced by an ALP field, the rotation angle becomes time-dependent. The net rotation angle for a photon observed today and emitted at conformal time $\eta$ is given by
\al{
    \beta(\eta) 
    &= \frac{g}{2}[\phi(\eta_0)-\phi(\eta)] 
    = \frac{g\phi_{\rm ini}}{2}\frac{\phi(\eta_0)-\phi(\eta)}{\phi_{\rm ini}}
    \,, \label{Eq:beta}
}
where $\eta_0$ is the present conformal time and $\phi_{\rm ini}$ is the initial value of $\phi$. 
To compute the impact of this time-dependent rotation on the CMB polarization, we solve the Boltzmann equation for the polarized photon distribution~\cite{Liu:2006:biref-time-evolve,Finelli:2009,Gubitosi:2014:biref-time,Lee:2013:biref}:
\begin{align}
    &_{\pm2}\Delta'_P + \iu q\nu~_{\pm2}\Delta_P 
    \notag \\
    &\qquad = a n_{\rm e}\sigma_T
        \left[
            -~_{\pm2}\Delta_P + \sqrt{\frac{6\pi}{5}}~_{\pm2}Y_{20}(\nu)\Pi(\eta,q)
        \right]
    \notag \\
    &\qquad\qquad \pm \iu g\phi'~_{\pm2}\Delta_P
    \,, \label{Eq:Boltzmann}
\end{align}
where $_{\pm2}\Delta_P$ are the Fourier modes of the linear polarization $Q \pm \iu U$ and are functions of conformal time $\eta$, wavevector magnitude $q$, and angle cosine between line-of-sight direction and wavevector $\nu$. The other symbols are: $a$ (scale factor), $n_{\rm e}$ (electron number density), $\sigma_{\rm T}$ (Thomson scattering cross-section), $\Pi$ (polarization source term)~\cite{Zaldarriaga:1996xe}, and ${}_\pm2 Y_{20}$ (spin-two spherical harmonics). Primes denote derivatives with respect to conformal time. 
The evolution of $\phi$ is governed by
\begin{equation}
    \phi'' + 2\frac{a'}{a}\phi' + a^2m_\phi^2 \phi = 0
    \,, \label{Eq:phi-EoM}
\end{equation}
assuming a quadratic potential $V(\phi)=m_\phi^2\phi^2/2$.
To derive the angular power spectra, $_{\pm2}\Delta_P$ is expanded as~\cite{Zaldarriaga:1996xe}:
\begin{align}
    _{\pm2}\Delta_P(\eta_0,q,\nu) 
    &\equiv -\sum_\l \iu^{-\l}\sqrt{4\pi(2\l+1)} 
    \notag \notag \\
    &\quad\times [\Delta_{E,\l}(q)\pm \iu\Delta_{B,\l}(q)] {}_{\pm2}Y_{\l0}(\nu) 
    \,, 
\end{align}
where the formal solution for the $E$- and $B$-mode sources is given as a line-of-sight integral \cite{Liu:2006:biref-time-evolve}: 
\begin{align}
    &\Delta_{E,\l}(q)\pm \iu\Delta_{B,\l}(q) = \frac{3}{4}\sqrt{\frac{(\l+2)!}{(\l-2)!}}
    \notag \\
    &\qquad\times \INT{}{\eta}{}{0}{\eta_0}an_{\rm e}\sigma_T \E^{-\tau(\eta)}\Pi \frac{j_l(x)}{x^2}\E^{\pm 2\iu \beta(\eta)}
    \,. 
\end{align}
Here, $j_\l(x)$ is the spherical Bessel function, $x=q(\eta_0-\eta)$, and $\tau(\eta)=\int_{\eta}^{\eta_0}{\rm d}\eta_1 an_{\rm e}\sigma_T$. Note that, the $E$-mode generated at time $\eta$ is multiplied by $\cos[2\beta(\eta)]$. Thus, a large variation of $\beta(\eta)$ during reionization leads to cancellations of signals among different epochs, resulting in a suppression of reionization signals in the $E$-mode power spectrum. The CMB angular power spectrum is then calculated as
\begin{equation}
    C_\l^{XY} = 4\pi
    \Int{}{(\ln q)}{} \mathcal{P}_s(q)\Delta_{X,\l}(q)\Delta_{Y,\l}(q)
    \,, \label{Eq:ClXY}
\end{equation}
where $X, Y \in {E, B}$ and $\mathcal{P}_s(q)$ is the dimensionless primordial scalar curvature power spectrum. Solving Eq.~\eqref{Eq:ClXY} yields the full shape of the birefringence-induced $EB$ power spectrum. We can systematically compute all CMB auto- and cross-power spectra by appropriately modifying the polarization transfer function, $\Delta_{X,\ell}$, within the {\tt CLASS} code \cite{CLASS}. 

We now incorporate the $n\pi$-phase ambiguity in the calculation of the CMB power spectra. The high-$\l$ Planck polarization data favor a constant rotation, and this is consistent with ALP masses below approximately $10^{-28}\,$eV, as inferred from the high-$\l$ Planck $EB$ power spectrum \cite{Namikawa:2025:Planck}. For ALP masses in the range $10^{-31}\,{\rm eV} \lesssim m_{\phi} \lesssim 10^{-30}\,{\rm eV}$, the ALP field evolves such that $\phi(\eta_0)=0$ today and $\phi(\eta)=\phi_{\rm ini}$ at recombination \cite{Nakatsuka:2022,Naokawa:2024xhn}. In this case, \eq{Eq:beta} reduces to $\beta_{\rm rec}=-g\phi_{\rm ini}/2$, where $\beta_{\rm rec}$ denotes the rotation angle at recombination. 
Recent measurements of the cosmic birefringence angle rely on the high-$\l$ $EB$ power spectrum, which is sourced predominantly at recombination. This implies that $\beta_{\rm rec}=\beta_0+180n\,{\rm deg}$, where $\beta_0=0.34\,$deg and $n$ is an integer reflecting the $n\pi$-ambiguity. 
Accordingly, for the mass range considered, the relation $-g\phi_{\rm ini}/2=\beta_0+180n,{\rm deg}$ holds. We use this relation to compute the CMB power spectra.

\begin{figure*}[th]
\bc
\includegraphics[width=8.5cm]{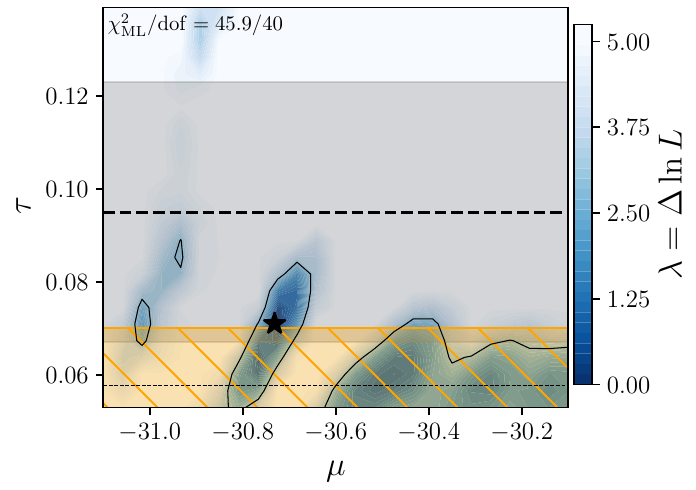}
\includegraphics[width=8.5cm]{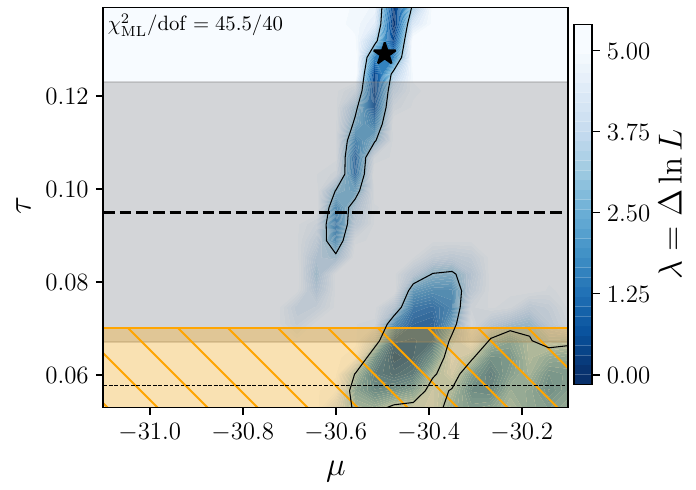}
\caption{
Likelihood bound on the ALP logarithmic mass, $\mu = \log_{10}(m_\phi, [{\rm eV}])$, and the optical depth, $\tau$, for a fixed value of $n = 1$ (left) and $n=2$ (right). We plot the log-likelihood difference with the $2\,\sigma$ confidence region (the black contour) that corresponds to $\lambda\equiv-\ln\mC{L}/\mC{L}_{\rm max}=3.0$. The constraints are obtained using the low-$\ell$ $EE + EB + BB$. The black stars are the maximum likelihood (ML) points. $\chi^2_{\rm ML}/{\rm d.o.f.}$ is the $\chi^2$ value at the ML points divided by the degree of freedom. The top gray bands, enclosed by black dashed lines, indicate the marginalized $2\,\sigma$ constraints on $\tau$ from Ref.~\cite{Sailer:2025:tau} derived without low-$\ell$ CMB data. The bottom thin black dashed line with the orange hatched region corresponds to the $2\,\sigma$ constraint on $\tau$ from Planck PR4~\cite{Tristram:2023:PR4:cosmo}.
}
\label{fig:const}
\ec
\end{figure*}

\begin{figure*}[t]
\bc
\includegraphics[width=8.5cm]{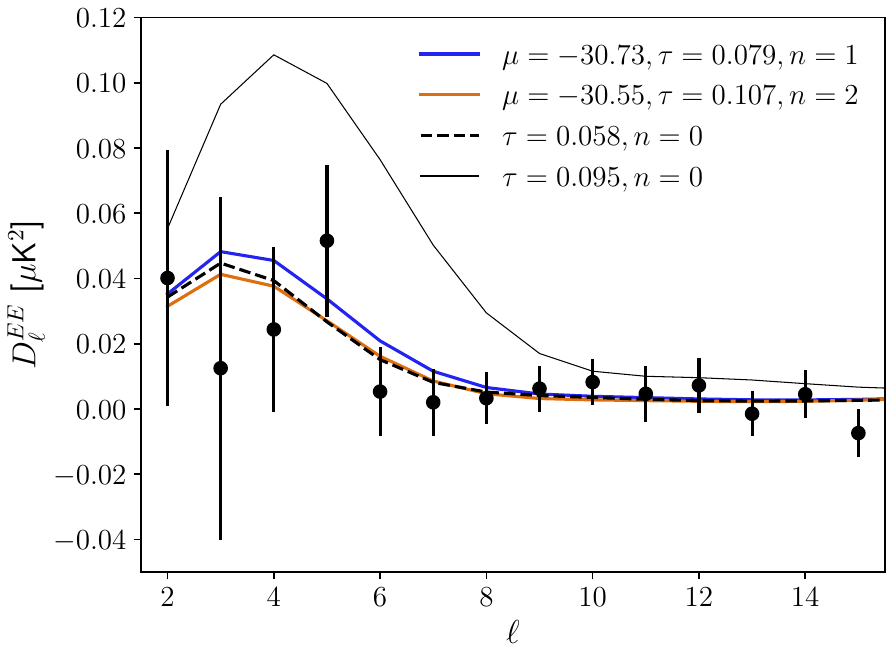}
\includegraphics[width=8.5cm]{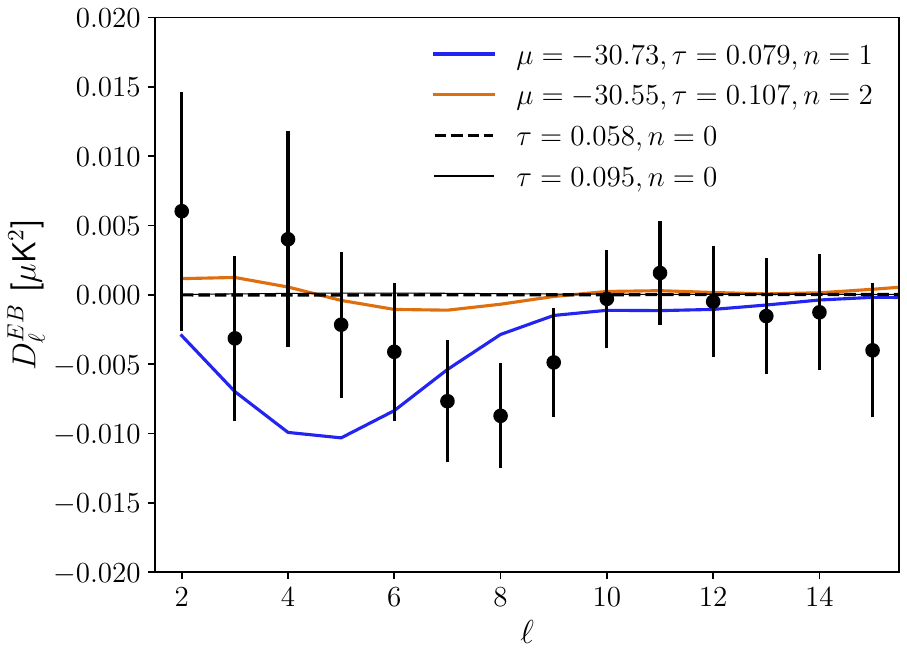}
\includegraphics[width=8.5cm]{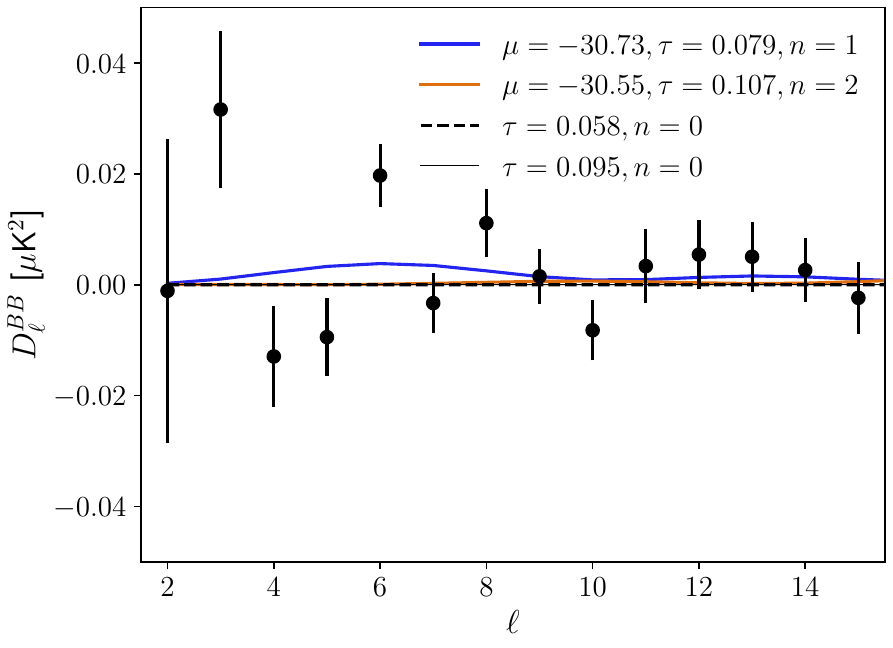}
\includegraphics[width=8.5cm]{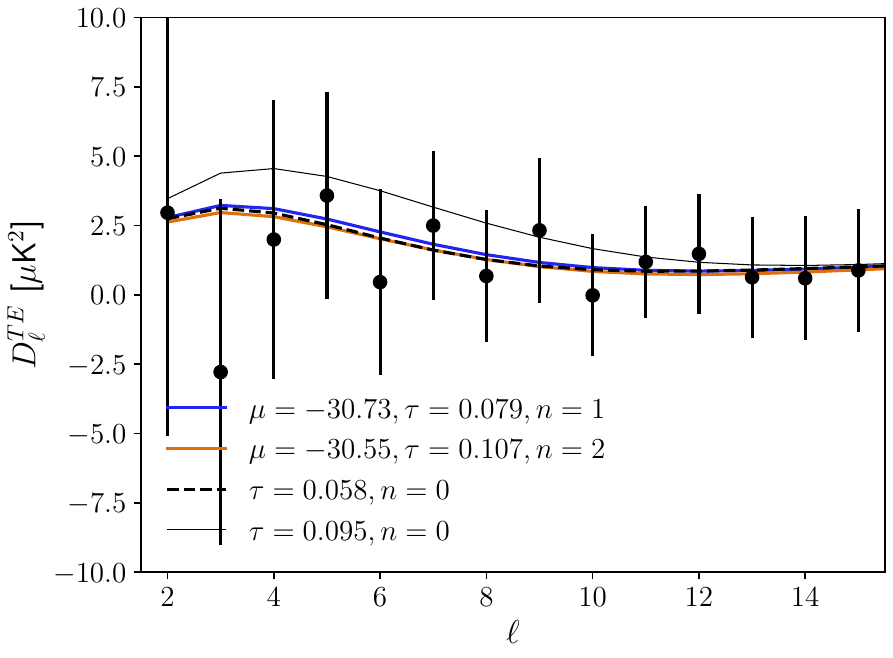}
\caption{
Angular power spectra of $EE$ (top left), $EB$ (top right), $BB$ (bottom left), and $TE$ (bottom right) for selected values of $\mu$, $\tau$, and $n$ that provide a good fit to both the low-$\ell$ polarization data and the optical depth constraint $\tau = 0.095 \pm 0.014$ (1$\sigma$). The black solid and dashed curves correspond to the theoretical spectra for $n=0$ with $\tau=0.095$ obtained without the low-$\l$ CMB data \cite{Sailer:2025:tau} and with $\tau=0.058$ obtained from the Planck PR4 analysis \cite{Tristram:2023:PR4:cosmo}, respectively. In these two curves, we fix $\mu=-31$ but the dependence of the spectra shown here on $\mu$ is negligible in the figures. The vertical axis shows $D_\l = \l(\l+1)C_\l/(2\pi)$. Observational data points for $EE$, $EB$, and $BB$ are taken from the Planck PR4 release~\cite{Tristram:2020:PR4:r}, while $TE$ data points are from Planck PR3~\cite{PR3:Cl}.
}
\label{fig:cls}
\ec
\end{figure*}

\ifdefined \submitletter
  {\it Parameter Search---}
\else
  \section{Parameter Search} \label{sec:results}
\fi
To quantitatively assess whether cosmic birefringence can account for the large value of $\tau$, we investigate the parameter space of $\mu\equiv\log_{10}(m_{\phi},[{\rm eV}])$, $\tau$, and $n$ that provides a good fit to the low-$\ell$ CMB polarization power spectra. In particular, we identify the regions of overlap with the observational constraint $\tau = 0.095 \pm 0.014$ ($1\,\sigma$). We do not consider variations around each integer $n$, as the resulting power spectra remain nearly unchanged---even when setting $\beta_0=0\,$deg. Thus, $n$ is treated as a discrete parameter.

We adopt the low-$\ell$ lollipop likelihood, $\mathcal{L}$, which includes the $EE$, $EB$, and $BB$ spectra~\cite{Tristram:2020:PR4:r}. Although cosmic birefringence also affects the $TE$ and $TB$ spectra, their modifications remain negligible compared to observational uncertainties within the parameter range of interest. We therefore focus exclusively on the low-$\ell$ polarization spectra. Among these, the $EE$ spectrum provides the dominant constraint on the parameters, while $EB$ carries some sensitivity in cases with $n = 1$. The $BB$ spectrum, in contrast, is largely insensitive to the parameter variations. The theoretical power spectra are computed using the best-fit cosmological parameters of the $\Lambda$CDM$+\sum m_\nu$ model from Ref.~\cite{Sailer:2025:tau}, which excludes the low-$\ell$ polarization data. 

We compute the log-likelihood difference $\lambda\equiv-\ln\mC{L}/\mC{L}_{\rm max}$ in the $\mu$-$\tau$ plane for fixed integer values of $n$. We evaluate the $2\,\sigma$ likelihood bound assuming that $\lambda$ follows an approximate chi-squared distribution with $2$ degrees of freedom. We also show the $\chi^2$ value at the maximum-likelihood point. The degree of freedom for the fitting is the total multipoles used in the $EE$, $EB$, and $BB$ spectra included in our analysis, minus the number of parameters used to fit. We also check that, within the parameter ranges of interest, modifications to the high-$\ell$ power spectra are negligible.

Figure~\ref{fig:const} presents the contour plots of the log-likelihood difference for $n=1$ and $n=2$. In both cases, there exist regions in the $\mu$–$\tau$ plane where the $2\,\sigma$ confidence region from the low-$\l$ polarization spectra overlaps with the $2\,\sigma$ constraints on $\tau$. We also examine the case of $n=3$, and similarly find the overlapped regions. As $n$ increases, the preferred value of $\mu$ that fits both the low-$\l$ polarization and the higher $\tau$ increases correspondingly. Note that the distinct regions in the $\mu$-$\tau$ plane originate from the nonmonotonic dependence of EB on $\mu$ and $\tau$. For example, certain $\mu$ values produce low-$\l$ EB features inconsistent with the data around $-30.9$ and $-30.7$ for $n=1$. For $n=2$ with $\mu\simeq -30.5$, an excluded range $0.07 \alt \tau \alt 0.11$ appears. As $\tau$ increases, reionization occurs earlier and the reionization-induced polarization is enhanced, causing the EB signal to be inconsistent with data. Beyond a certain threshold, however, the EB amplitude decreases because contributions from different reionization epochs partially cancel through line-of-sight integration of the rotated polarization, bringing the EB signal back into agreement with observations.

Figure~\ref{fig:cls} shows the low-$\l$ polarization power spectra for specific values of $\mu$ and $\tau$ taken from the overlapped regions. The power spectra for the $n=1$ and $n=2$ cases show good agreement with the data for specific combinations of $\mu$ and $\tau$. For comparison, we also show the spectra for $(n,\tau)=(0,0.095)$ and $(0,0.058)$ with $\mu=-31$. Note that the spectra are nearly insensitive to variations in $\mu$. The case $(n,\tau)=(0,0.095)$ yields $\chi^2/$d.o.f.$=88.5/40$, corresponding to a discrepancy greater than $5\,\sigma$ with the data, and thus clearly fails to reproduce the low-$\l$ polarization power spectra. Additionally, we present the $TE$ power spectrum, which was not used in the fitting procedure. The predicted $TE$ spectrum agrees well with the Planck data. We also check the low-$\l$ $TB$ spectrum obtained from Planck PR2 data \cite{PR2:Cl} and confirm that the theoretical predictions remain well within the observational uncertainties, owing to the large noise level in the $TB$ measurements. 

\ifdefined \submitletter
  {\it Summary and Discussion---}
\else
  \section{Summary and Discussion} \label{sec:discussion}
\fi
We pointed out that cosmic birefringence helps to resolve the tension in the neutrino mass inferred from CMB and DESI BAO observations if the birefringence angle varies significantly during reionization. Specifically, we have demonstrated that cosmic birefringence with a nonzero integer $n$ can simultaneously explain the recent Planck $EB$ power spectrum signal and a large value of $\tau \simeq 0.09$, which helps alleviate the tension. 

The constraint on the cosmic birefringence rotation angle can be translated into a bound on the photon-ALP coupling constant $g$. Based on the results in Refs.~\cite{Fujita:2020:forecast,Fujita:2020:biref}, a rotation angle of $\beta_0=0.34\,$deg corresponds to $g \gtrsim 1.5 \times 10^{-20}$ GeV$^{-1}$ for the $\mu$ range relevant to this work, where the lower bound arises from large-scale structure constraints on ALP energy density. For nonzero $n$, the coupling constant increases by a factor of approximately $180|n|/0.34 \simeq 5.2 \times 10^2$ \cite{Naokawa:2024xhn}. For example, for $n=2$, we obtain $g \gtrsim 2.8 \times 10^{-17}$ GeV$^{-1}$, which is still well below the upper bound from Chandra X-ray observations ($g \lesssim 10^{-12}$ GeV$^{-1}$) \cite{Berg:2016:axion}.

We have considered the case of a large rotation angle, which requires a large value of $|g\phi_{\rm ini}/2|$. Writing $g=\alpha/(4\pi^2 f_\phi)$, with $\alpha \simeq 1/137$ denoting the QED fine-structure constant and $f_\phi$ the decay constant, we obtain $g\phi_{\rm ini}/2=(\alpha/4\pi)(\phi_{\rm ini}/2\pi f_\phi)$. To realize $g\phi_{\rm ini}/2 \simeq \pi$, one needs $\phi_{\rm ini}/(2\pi f_\phi) \sim 4\pi^2/\alpha \simeq 5.4 \times 10^3$. This shows that such a large rotation angle cannot be realized if $\phi_{\rm ini}$ is restricted to a compact field space of circumference $2\pi f_\phi$. Several mechanisms, such as the clockwork mechanism and kinetic mixing \cite{Babu:1994:kinetick-mixing,Choi:2015:clockwork,Kaplan:2015:clockwork,Farina:2016:clockwork,Agrawal:2018:axion}, have been proposed to realize an enhanced photon-ALP coupling, which could in turn produce a large rotation angle. A detailed exploration of such models in the context of cosmic birefringence has not yet been fully developed and remains an interesting avenue for future work.

In this study, we assumed a single ALP with a quadratic potential to account for cosmic birefringence. However, multiple ALPs may also contribute to the effect. For instance, Ref.~\cite{Namikawa:2023:pSZ} considers a two-ALP model in which one ALP begins to oscillate between recombination and reionization, while the other oscillates after reionization. Denoting the rotation angles induced by each ALP as $\beta_1$ and $\beta_2$, the total rotation angle becomes $\beta_1 + \beta_2$. In this scenario, the rotation angle at recombination is $\beta_{\rm rec} = \beta_1 + \beta_2$, while at reionization it is approximately $\beta_{\rm rei} \simeq \beta_2$. If the two ALPs have opposite-sign initial field values with similar magnitudes and a large photon-ALP coupling, the two contributions can nearly cancel, yielding $\beta_{\rm rec} \simeq 0.34\,$deg and $\beta_{\rm rei} \sim \mathcal{O}(180\,{\rm deg}) \gg \beta_{\rm rec}$. This mimics the observational effect of the $n\pi$-phase ambiguity. More generally, any model that realizes $\beta_{\rm rec} \simeq 0.34\,$deg and $\beta_{\rm rei} \sim \mathcal{O}(180\,{\rm deg})$ would produce similar polarization spectra.

While we focused on the preference of negative neutrino mass arising from the mild tension between DESI BAO and CMB, other cosmological probes can yield slightly different constraints on $\Lambda$CDM parameters. For example, combining CMB with Type Ia supernova luminosity distances leads to a smaller best-fit value of $\tau$ \cite{Jhaveri:2025:tau}. Recently, Ref.~\cite{Cain:2025:pkSZ} pointed out that the patchy kinematic Sunyaev–Zel’dovich (pkSZ) effect, measured by the South Pole Telescope, may be in tension with a large $\tau$. However, pkSZ constraints are derived from high-$\ell$ power spectra and may be affected by systematic uncertainties from extra-Galactic sources. Moreover, the amplitude of the pkSZ signal depends sensitively on the underlying reionization model.

Future measurements of large-scale $EE$, $EB$, and $BB$ polarization spectra by Cosmology Large Angular Scale Surveyor \cite{CLASS:2025} and LiteBIRD \cite{LiteBIRD,LiteBIRD:2025:biref} will offer a stringent test of the cosmic birefringence scenario proposed here. Additional late-time observables can also provide complementary constraints. For example, radio galaxies have been used to probe birefringence, although current analyses are limited by uncertainties in the intrinsic polarization directions due to a limited number of sources \cite{Cimatti:1994:rotation,Carroll:1997:radio,Naokawa:2025shr}. Other promising methods include using galaxy shapes~\cite{Yin:2024:galaxy} and the polarized Sunyaev–Zel’dovich effect~\cite{Hotinli:2022:pkSZ,Lee:2022:pSZ-biref,Namikawa:2023:pSZ}. A birefringence signal consistent with our constraint implies a time-dependent rotation angle at low redshift. Therefore, by observing large samples of low-redshift objects with well-characterized intrinsic polarization, we may be able to constrain the redshift dependence of the birefringence angle and further test the parameter space identified in this work.


{\it Acknowledgement}---We thank Gerrit Farren for providing the best-fit values of cosmological parameters in Ref.~\cite{Sailer:2025:tau}. We also thank Blake Sherwin and Fumihiro Naokawa for useful discussion. We acknowledge support from JSPS KAKENHI Grant No. JP20H05859, No. JP22K03682, No. JP24KK0248, and No. JP25K00996. The Kavli IPMU is supported by World Premier International Research Center Initiative (WPI Initiative), MEXT, Japan. This work uses resources of the National Energy Research Scientific Computing Center (NERSC). 

{\it Data availability}---The data that support the findings of this article are not publicly available. The data are available from the authors upon reasonable request. 




\bibliographystyle{mybst}
\bibliography{cite}

\end{document}